\documentclass[%
 reprint,
superscriptaddress,
 amsmath,amssymb,
 aps,
]{revtex4-2}

\usepackage{graphicx}%
\usepackage{dcolumn}%
\usepackage{bm}%

\begin{document}

\preprint{APS/123-QED}

\title{Data is often loadable in short depth: Quantum circuits from tensor networks for finance, images, fluids, and proteins}%

\author{Raghav Jumade}
\affiliation{Intel Corporation, Hillsboro, OR 97124, USA}
\author{Nicolas PD Sawaya}
\email{nsawaya@protonmail.com}
\affiliation{Intel Labs, Santa Clara, CA 95054, USA (former affiliation)}
\affiliation{HPI Biosciences, Oakland, CA 94608, USA; \textit{and} Azulene Labs Inc., San Francisco, CA 94115, USA (current affiliations)}

\begin{abstract}
Though there has been substantial progress in developing quantum algorithms to study classical datasets, the cost of simply \textit{loading} classical data is an obstacle to quantum advantage. When the amplitude encoding is used, loading an arbitrary classical vector requires up to exponential circuit depths with respect to the number of qubits. Here, we address this ``input problem'' with two contributions. First, we introduce a circuit compilation method based on tensor network (TN) theory. Our method---AMLET (Automatic Multi-layer Loader Exploiting TNs)---proceeds via careful construction of a specific TN topology and 
can be tailored to arbitrary circuit depths. 
Second, we perform numerical experiments on real-world classical data from four distinct areas: finance, images, fluid mechanics, and proteins. To the best of our knowledge, this is the broadest numerical analysis to date of loading classical data into a quantum computer. The required circuit depths are often several orders of magnitude lower than the exponentially-scaling general loading algorithm would require. Besides introducing a more efficient loading algorithm, this work demonstrates that many classical datasets are loadable in depths that are much shorter than previously expected, which has positive implications for speeding up classical workloads on quantum computers.
\end{abstract}

\maketitle

\textbf{Introduction. }
Though the most natural application area for quantum computation is the simulation of quantum physics, there has been substantial recent theoretical progress in quantum algorithms for classical problems. For instance, algorithms have been developed for solving linear systems \cite{harrow2009quantum}, simulating differential equations \cite{berry2014ode,lloyd2020pde,lubasch2020variational,childs2021pde}, %
machine learning \cite{perdomo2018qml,schuld2019qml,cerezo2022qml}, and various tasks in finance \cite{herman2023quantum} and image processing \cite{zhang2015qsobel,jiang2019improved,jiang2015quantum,li2018multi,caraiman2013quantum,caraiman2014histogram,nakaji2021quantum}. 

However, it is often the case that simply \textit{loading} classical data into the quantum computer is so costly that it overwhelms any quantum advantage that would have been expected in the execution of the main algorithm \cite{aaronson2015fineprint,tang21qpca}. In the general case, exact state preparation for a vector of length $M=2^n$ into $n$ qubits requires a circuit depth exponential in $n$ \cite{aaronson2015fineprint,plesch2011univgatedecomp,poulin2011quantum} if ancilla qubits are not used. 
This ``input problem'' suggests two distinct research directions. First, it may be the case that some industrially relevant classical data, which is often quite structured, can in fact be loaded efficiently---hence studying existing classical data would bring practical insights and allow us to identify such data classes. Second, this problem motivates the development of algorithms that are more efficient for loading data, in order to mitigate this bottleneck.

In this work, we contribute to each of these research directions. First we introduce a novel tensor network (TN) algorithm called AMLET (Automatic Multi-layer Loader Exploiting TNs), demonstrating its improved performance relative to a modified version of a previously described TN algorithm for loading matrix product states from quantum problems. Second, we perform a broad numerical study to determine required circuit depths for loading various types of real-world classical data, including images, turbulent fluid data, financial data, and protein configurations.

There are several previous methods that attempt to address the input problem \cite{giovannetti2008qram,plesch2011univgatedecomp,zoufal2019gan,araujo2021divide,korzekwa2022classicalinfo,bausch2022blackbox,larose2020robust,holmes2020stateprep,marin2021loading,shirakawa2021automatic,zhang21lowdepth,amankwah2022quantum,camps2022fable,rudolph2022mps,west2023drastic,melnikov2023statepreptn,akhalwaya2023modular}; %
we highlight some prominent approaches here. There are general methods with fixed circuit structures that scale exponentially with the number of qubits \cite{plesch2011univgatedecomp}; such methods are not easily amenable to tailorable space-depth tradeoffs. There are methods that can accurately load only some classes of data, for example smooth differentiable functions \cite{holmes2020stateprep}. Others have proposed using neural network methods to generate quantum circuits for loading data \cite{zoufal2019gan}; such methods must be trained on the data class of interest and there does not appear to be a straightforward way to control error. Another notable strategy is to reduce depth at the cost of introducing additional ancilla qubits \cite{zhang21lowdepth,gui2023spacetime}. Relatedly, quantum random-access memory (QRAM) \cite{giovannetti2008qram} promises to partially mitigate the input problem, though such a solution would require more complex hardware and substantially more qubits. In this work, we show that AMLET produces accurate, optimization-free, ancilla-free, short-depth state preparation circuits for the amplitude encoding.

\begin{figure}
    \centering
    \includegraphics[width=\linewidth]{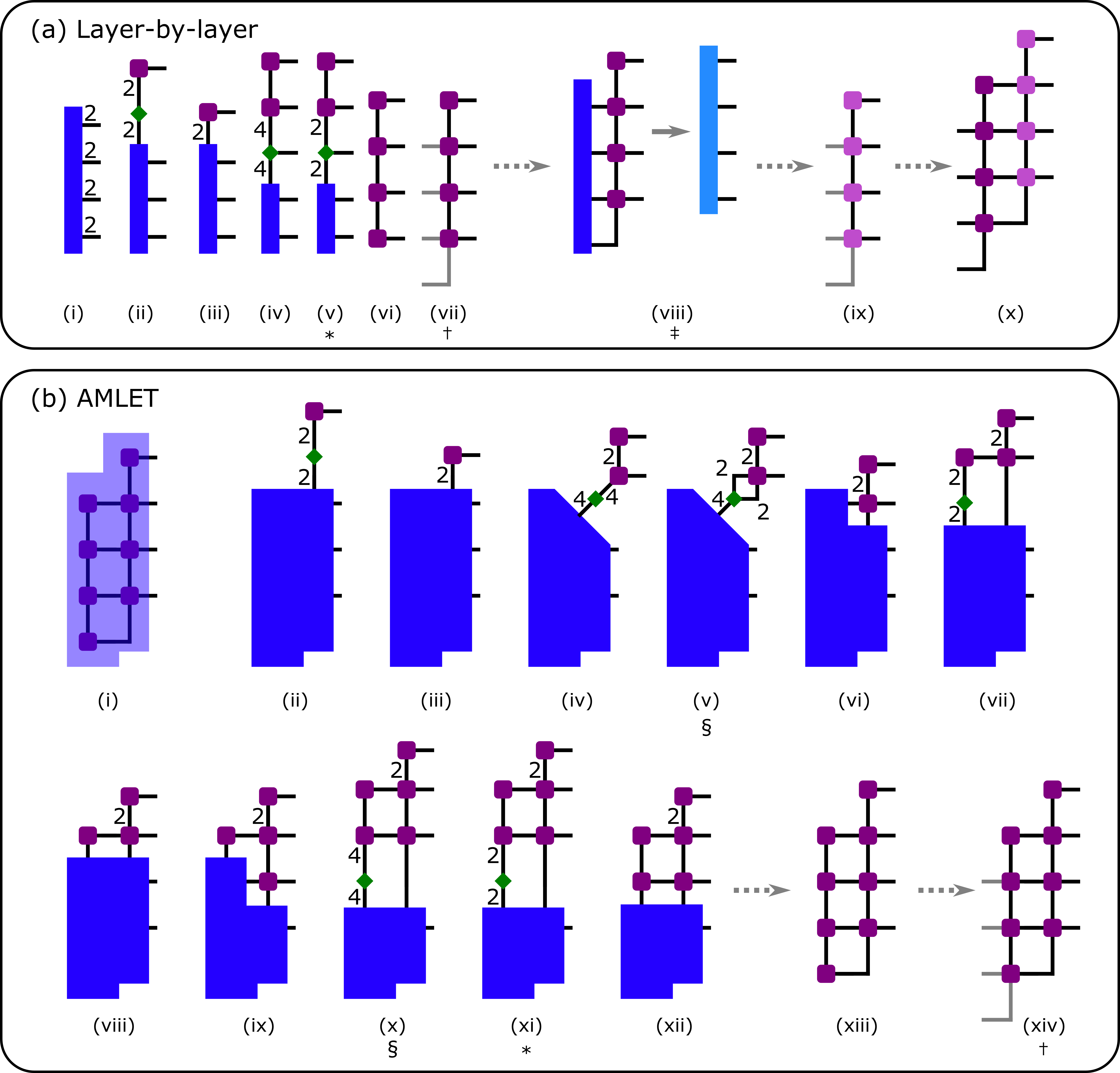}
    \caption{Schematics of the 2-layer versions of the (a) layer-by-layer and (b) AMLET algorithms. Both algorithms may be generalized to an arbitrary number of layers $\mathcal D$. Green triangles denote the diagonal singular value tensor that results from performing SVD. `*' indicates that SVD truncation was performed, `$\dag$' that tensors were expanded to form unitary quantum gates, `$\ddag$' that the full circuit was contracted, and `$\S$' that reindexing was performed. All singular value tensors (green) must be contracted ``downwards,'' such that the unitary-like normalization condition is always conserved above. Notably, while the first approximation (SVD truncation) in layer-by-layer is at step \textit{a.v}, two-layer AMLET does not require any approximation until step \textit{b.xi}; more AMLET layers would result in delaying the first truncations even further and in fewer total truncations.}
    \label{fig:schem}
\end{figure}

\textbf{Algorithms for loading data.}
We begin with a high-level description of both circuit compilation algorithms. The first algorithm we implement is somewhat similar to the work of Ran \cite{ran2020encoding}, originally proposed not for classical data but for loading a matrix product state (MPS). We refer to it as the layer-by-layer algorithm. First one approximates the $2^n$-length vector as an MPS, via repeatedly performing singular value decompositions (SVD) and truncations \cite{orus2014practical}. Then indices (\textit{i.e.} edges) are added to the tensors on the edge of the tensor network to form two-qubit gates (rank-four tensors), as shown schematically in Figure \ref{fig:schem}, where SO(4) symmetry is enforced by populating the new tensor values with the null space of the original lower-dimensional tensor. Note that a gate in SO(4) may be decomposed into an arbitrary gate set \cite{vatan2004so4}. (Throughout this work we assume a gateset of CNOTs and arbitrary one-qubit rotations.) %
One then contracts this ``disentangling circuit'' with the state vector, yielding a new $2^n$-length vector. The full procedure is then repeated on the resulting vector, and the final quantum circuit is the concatenation of the resulting disentangling circuits.

Next we introduce our AMLET algorithm, which instead produces the entire quantum circuit ``on the fly.'' SVD-and-truncation decompositions form the basis of this algorithm as well, but the important difference is that we split each new bond into two bonds each of size 2, while retaining the desired network structure. AMLET is more complex, requiring careful fusing, splitting, and reordering of indices, while keeping track of tensor positions.

Both methods allow for a monotonic trade off between accuracy and depth---if one needs to increase the accuracy of the loading circuit, one can do so simply by successively increasing the depth. Additionally, neither requires the optimization of circuit parameters (which can lead to pathologies such as barren plateaus \cite{mcclean2018barren}), as the constructed quantum circuit results directly from successive linear algebra steps in a deterministic number of operations.
All algorithms were implemented using in-house code built on Scipy \cite{scipy} and Quimb \cite{quimb}. %

Schematic descriptions of the layer-by-layer and AMLET algorithms are shown in Figure \ref{fig:schem}. Each begins with a normalized $2^n$-by-1 vector. As mentioned, the layer-by-layer algorithm is conceptually similar to previous work \cite{ran2020encoding, rudolph2022mps} that used an SVD-and-truncation method to load a matrix product state (MPS) describing a physical system. The green diamonds are the diagonal singular value (SV) tensors, which result from SVD. For both algorithms, the green diamond is always pushed ``downwards," which ensures that the ``above" tensors retain the left-normalized form \cite{schollwock2011dmrg}, i.e. the orthogonalization condition 
\begin{equation}
\sum_{\sigma,i} U_{i;i-1,\sigma}^\dag U_{i-1,\sigma;i'} = I_{i,i'}
\end{equation}
where $i$ and $i+1$ are respectively the indices above and below the tensor, $I$ is the identity operator, and $\sigma$ is one of the original indices. Another required routine is SVD truncation, whereby the middle index of the tensor decomposition is reduced to dimension $K$,
\begin{equation}
\sum_{a,b,c,d}^{l_a,l_b,l_c,l_d} U_{ab} S_{bc} V_{cd}  \rightarrow  \sum_{a,b,c,d}^{l_a,K,K,l_d} U_{ab} S_{bc} V_{cd}
\end{equation}
where we have assumed singular value matrix $S$ is ordered from largest to smallest value. For example in steps \textit{a.v} and \textit{b.xi} the tensors must be truncated to $K=2$, to ensure that the eventual tensor network maps directly to rank-4 (\textit{i.e.} two-qubit) tensors of appropriate size.

In steps \textit{a.vii} and \textit{b.xiv}, tensors in the network are expanded (\textit{i.e.} indices are added) in order to create two-qubit gates. This tensor expansion results in gauge freedom, whereby additional tensor elements must be populated. Expanding a $2 \times 2 \times 2$ to a $2 \times 2 \times 2 \times 2$ tensor, and rearranging the elements to correspond to a $4 \times 4$ two-qubit quantum gate, we schematically describe this expansion via 
\begin{equation}
\begin{split}
\begin{pmatrix}
* & * & \cdot & \cdot \\
* & * & \cdot & \cdot \\
* & * & \cdot & \cdot \\
* & * & \cdot & \cdot \\
\end{pmatrix}
\rightarrow
\begin{pmatrix}
* & * & * & * \\
* & * & * & * \\
* & * & * & * \\
* & * & * & * \\
\end{pmatrix}
\end{split}
\end{equation}
where $*$ denotes already-filled entries and $\cdot$ denotes undetermined entries (\textit{i.e.} gauge freedom). Determining the empty columns \cite{ran2020encoding} is equivalent to finding the null space of the filled columns, which may be done using a Gram-Schmidt type of procedure. Note that the bottom-most tensors in Figure \ref{fig:schem} instead require that three columns be filled.
Finally, AMLET requires the additional subroutine of index splitting (reshaping) \cite{singh2011u1symm},
\begin{equation}
M_{ab} \rightarrow \tilde M_{acd}; \; \; \; b = c \times d
\end{equation}
which requires careful realignment of the tensor network to ensure correct ordering of tensor entries. This step is shown for example in \textit{b.v}. Notably, the use of splitting in step \textit{b.v} instead of truncating as in \textit{a.v} ensures that approximations in AMLET are introduced significantly later, which in turns leads to fewer approximation (\textit{i.e.} truncation) steps overall. By contrast, the layer-by-layer algorithm requires the introduction of approximations after almost every SVD operation. This is likely the reason for AMLET out-performing layer-by-layer in our results below.

The classical complexity of constructing the quantum circuit scales at most as $O(M)$ if the depth is bounded by a constant, by the following reasoning. Stated differently, forming a $\log_2(M)$-tensor MPS from a $M$-length vector scales at most as $O(M)$ if naive matrix subroutines are used. 
(More advanced subroutines \textit{e.g.} the known $O(Q^{2.373})$-scaling matrix-matrix multiplication algorithm for $Q \times Q$ matrix can reduce this asymptotic scaling \cite{burgisser2013algebraic}.) This is because naive SVD requires $O(ab^2)$ steps for an $a \times b$ matrix, where the first SVD-and-truncation is of a $2 \times \frac{M}{2}$ matrix ($ab^2=2M$), the second is of a $2 \times \frac{M}{4}$ matrix ($ab^2=M$), and so on. 
In turn this leads to the sum $\sum_i^n \frac{M}{2^i} \sim O(M)$. When the number of layers is bounded by a constant a similar argument yields $O(M)$ for the AMLET algorithm, albeit with a different prefactor. 

\begin{figure}
\centering
\includegraphics[width=\linewidth]{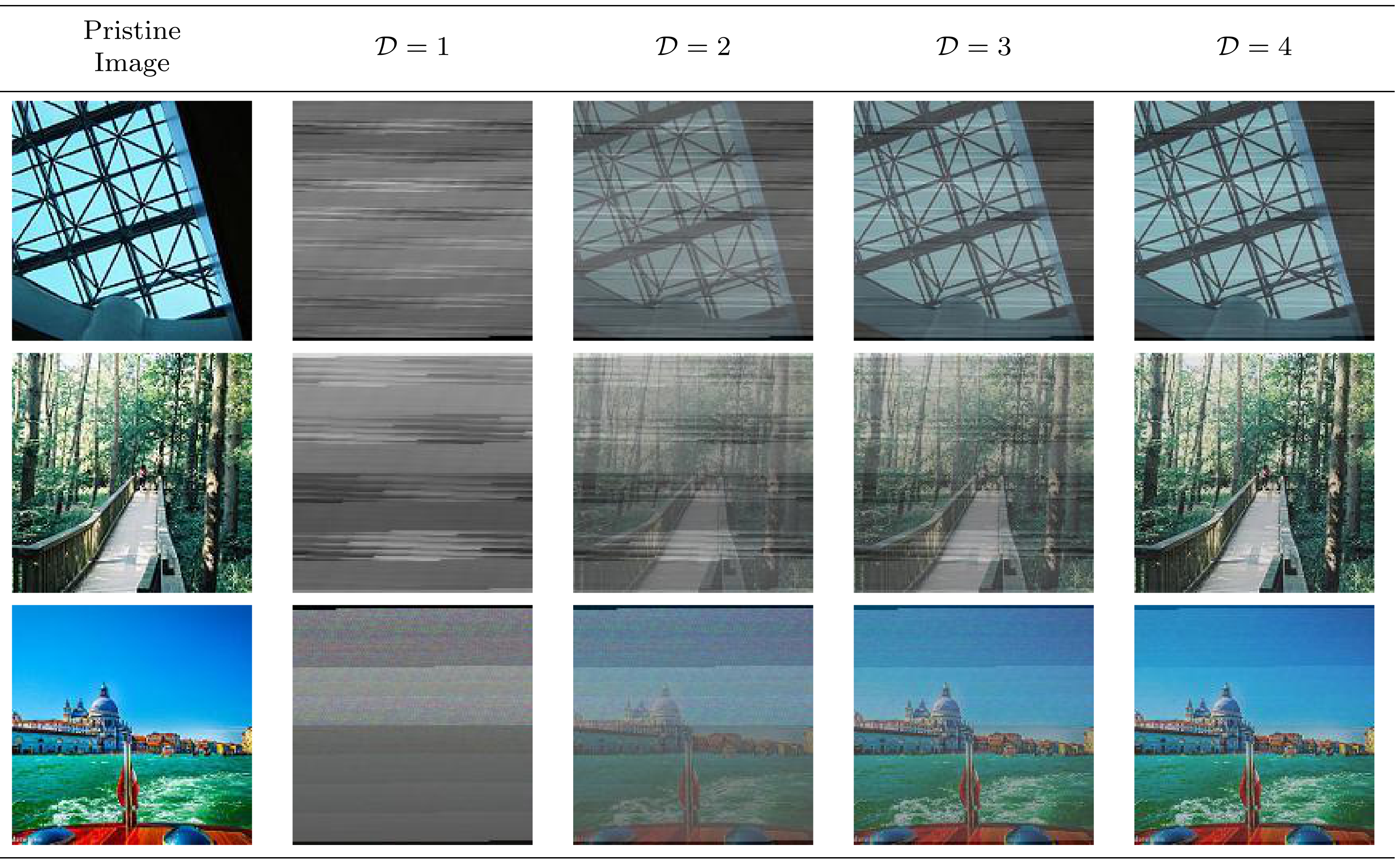}
\caption{Resulting approximate images from 14-qubit AMLET quantum circuits for $\mathcal{D}=$ 1 through 4. In order the demonstrate that even the poorest results were relatively short-depth, these three images were chosen because they yielded the lowest fidelity in their image class.}
\label{fig:2}
\end{figure}

\textbf{Classical datasets.}
Here we summarize the four types of classical data used in this work. Each uses a slightly different application-motivated quality measure. Note that in all cases the measure is based on the \textit{unnormalized} vector, meaning that an increase in the vector norm of the original state does indeed lead to a more difficult loading problem. If we were to take the (flawed) approach of considering quality measures on normalized vectors, the circuit depths would be substantially shorter.

\textit{Images. }
We use the Intel Image Classification dataset \cite{intel-images}, which contains six categories (buildings, forests, \textit{etc.}) and was originally designed to benchmark classification algorithms \cite{wu2020using-intel}. We encode the two-dimensional RGB (red-green-blue) images in row-major format, where each color is grouped together (\textit{i.e.} all red values are listed before all blue values). Note that researchers have proposed several other more complex formats \cite{yan2016imagereprsurvey} to represent images in quantum computers. The quality metric we use is
\begin{equation}\label{eq:vecdiff2norm}
\| \vec v_{exact} - \vec v_{approx} \|_2 \leq \epsilon_I
\end{equation}
where for images we set an arbitrary threshold $\epsilon_I < 10^{-3}$ where all vectors were length-$2^{14}$. %

\textit{Fluid dynamics.} Accurately simulating the Navier-Stokes equations in the turbulent regime is computationally expensive and often intractable for real-world problems \cite{rodi2017turbulence}, and there has been some effort in quantum algorithm development in this area \cite{lloyd2020pde}. We expect velocity data for laminar flow to be significantly easier to load than data from turbulent data, as the latter is less smooth. Hence we focus on (incompressible) turbulent velocity data (\textit{i.e.} high Reynolds number) from a recently published dataset 
\cite{mcconkey2021fluids}, where we limit ourselves to encoding only $U_x$, the velocity vector along the direction of the bulk flow. 
For this dataset, we ensure that the loaded data meets two error conditions. First, we bound $\epsilon_I$ by a constant. Second we ensure that total momentum is conserved in the incompressible fluid to within some error, which is achieved by ensuring that
\begin{equation}\label{eq:fluidmom}
|\sum(v_{exact}) - \sum(v_{approximate})| < \epsilon_M
\end{equation}
where this expression may be derived by summing the velocities to produce momentum, and then bounding the error in momentum $|M_{exact}-M_{approx}|$. %
We estimate that choosing $\epsilon_M<100$ yields a maximum relative error of at most $10^{-4}$ in the momentum of the entire block of fluid.

\textit{Finance.} We use a dataset of minute-by-minute data for the following stock indices: S\&P 500,  NIKKEI 225, DAX 30, and EUROSTOXX \cite{finance-kaggle-2020}. We use equation \eqref{eq:vecdiff2norm} as a quality metric, with $\epsilon_I$ = 1 currency unit (dollar, Euro, etc.). Notably, because the data is recorded only during trading hours, there are many sudden changes in value when two adjacent time points come from different days, an additional roughness that likely makes accurate loading more difficult.

\textit{Protein atomic positions.} Proteins are a versatile and ubiquitous class of biomolecule, commonly studied by direct physics simulation \cite{sledz2018proteinmd} and also via statistical and machine learning models \cite{baek2021uwbaker,jumper2021alphafold}. Quantum algorithms have been proposed for porting such workloads to quantum computers \cite{robert2021proteinqalg,casares2022qfold}. The metric we use is the atomic root mean square deviation (RMSD), 
\begin{equation}
\sqrt{ \frac{1}{N_a} \sum_i^{N_a} \| \vec r_{j,exact} - \vec r_{j,approx} \|^2 } < \epsilon_A,
\end{equation}
where $\vec r_i$ is the three-coordinate position of atom $j$ of $N_a$ total atoms. This is proportional to equation \eqref{eq:vecdiff2norm} with an $1/M$ multiplicative factor. We use $\epsilon_I < 1 $ \AA, as in the vast majority of cases this is more precise than X-ray crystallography can achieve. 
We chose three of the largest proteins available in the Protein Data Bank \cite{burley2021rcsbpdb}: protein entries 1VVJ \cite{maehigashi2014_1VVJ}, 3J3Q \cite{zhao2013_3J3Q}, and 3JA1 \cite{li2015_3JA1} are related to ribosome function, the HIV-1 capsid, and protein translation, respectively. Note that the data includes some atoms from RNA molecules. To obtain a larger dataset for each qubit count, we load non-overlapping sequences of atoms (hence the number of samples decreases as the qubit count increases).

\begin{figure}
    \centering
    \includegraphics[width=\linewidth]{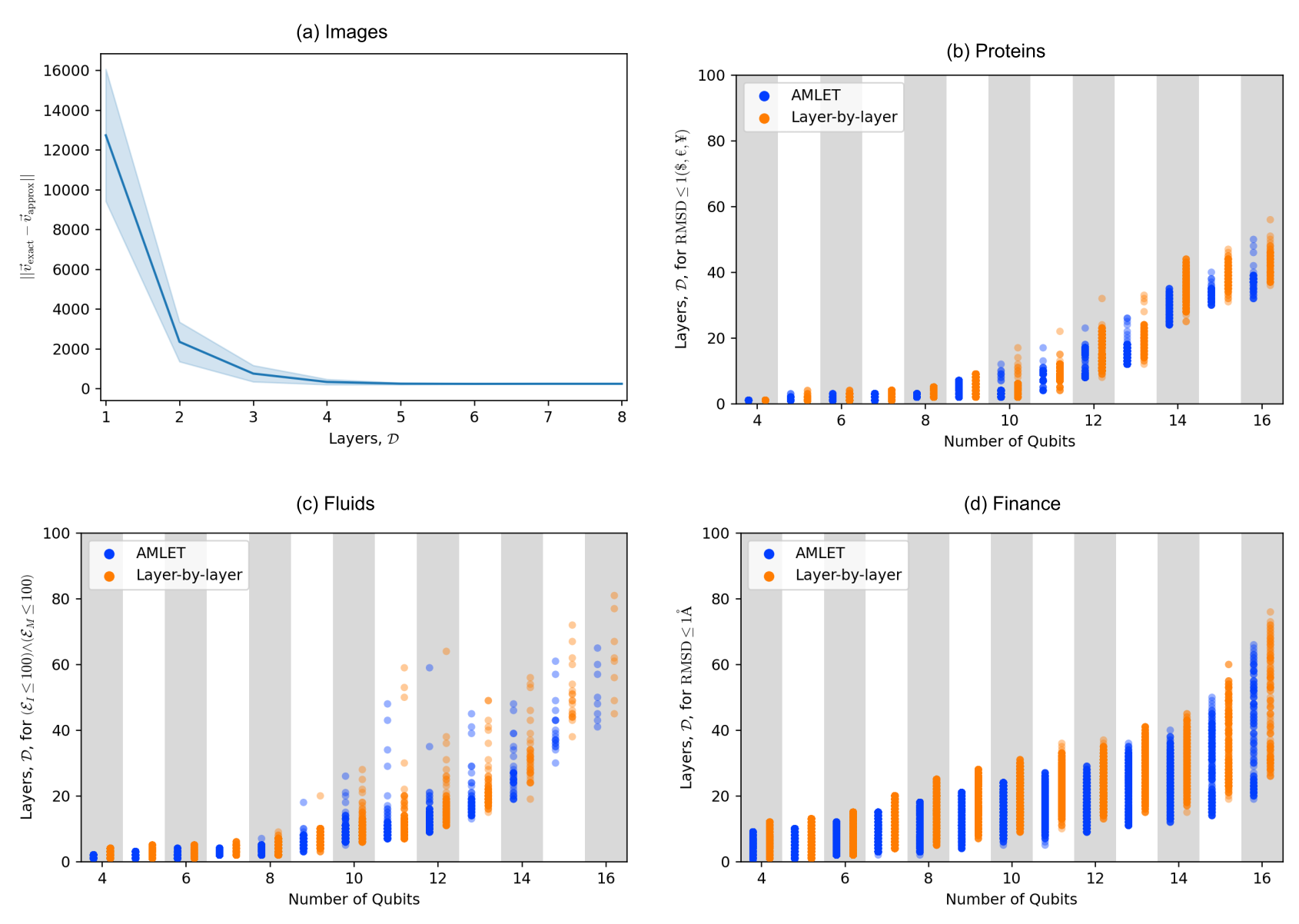}
    \caption{Plot (a) shows the reduction in error in various images with increase in number of layers, using the AMLET algorithm; the plotted data is bounded by the maximum and minimum values over all images in the dataset. Plots (b), (c) and (d) show the number of layers needed to meet the given accuracy criteria for vectors representing positions of protein atomic coordinates, fluid velocity, and financial market indices respectively. The size of each classical vector is $2^n$. 
Importantly, for all four classical application areas, the circuit depths are far shallower than the general amplitude encoding loading algorithm (a general amplitude encoding circuit would require the equivalent of at least $\mathcal D=215$ layers for 12 qubits and of $\mathcal D=2,715$ layers for 16 qubits, much larger than the vertical plot limits). Each of the application-specific error criteria are defined in the text.}
    \label{fig:3}
\end{figure}

\textbf{Results \& Discussion.}
We begin with the most important obervation from our results: though the depths are substantial in a subset of cases the depth does not increase exponentially, even though exactly loading the most general state indeed requires exponential depth \cite{aaronson2015fineprint, plesch2011univgatedecomp, poulin2011quantum}. A general amplitude encoding circuit \cite{plesch2011univgatedecomp,vatan2004so4} would have required the equivalent of at least $\mathcal D=215$ layers for 12 qubits and of $\mathcal D=$2,715 layers for 16 qubits, while $\mathcal D < 90$ for AMLET (up to 16 qubits) with respect to all of our data sets.

The image dataset is the easiest to load of these data classes, likely because it is the most structured. Figure \ref{fig:2} visually demonstrates how quickly the features of the image become clear. After four layers, virtually all features are present. Note that the image dataset was the only set that was studied in an ``intensive" as opposed to ``extensive'' way---in other words, we compressed each image to 14 qubits. During our tests we observed that the depth does not change substantially if we use a larger (less compressed) representation of the image, likely because there is little information being removed during compression.

For protein data, the number of layers increases smoothly, and a large subset of atomic coordinates requires just $\mathcal D < 40$ even for 20,000 atoms. Note that 16 qubits corresponds to $2^{16}/3 \approx 21,800$ atoms. Titin, the largest known protein, with $\sim$345,000 atoms, would require just $\log_2 (3 \times 345,000) =$ 20 qubits to load. Hence if the trend in Figure \ref{fig:3}(b) persists then a quite conservative prediction is that even the largest known protein might be loadable in $\mathcal D < 200$.

The turbulent fluid data requires the highest circuit depths of the data we considered, though even 16-qubit vectors are loadable in a relatively few layers $\mathcal D < 90$. We attribute the anomolous jumps in maximum depth---\textit{e.g.} the worst case 12-qubit circuit is deeper than the worst-case 14-qubit circuit---to error cancellation in the momentum, \textit{i.e.} including more data points will often lead to a smaller results for formula \eqref{eq:fluidmom}.

Finally, the financial data shows a moderate increase in depth as qubit count increases, suggesting that financial time-series data is often loadable in manageable depths even up to tens of thousands of data points. Notably, the range (standard deviation) of depths is relatively small for each qubit count, suggesting that there is less variance in loading difficulty for financial data. This was not the case for the protein atomic positions, for which even some large vectors were loadable in very short depth. 

In all cases, AMLET yields shorter depths than the layer-by-layer method for the same error. At 16 qubits, the reduction in is $\mathcal D$ between 8\% and 25\%. We attribute this to far fewer SVD truncations being required in the AMLET algorithm.

\textbf{Conclusion.}
We have introduced AMLET, a tensor network-based compilation algorithm for loading a length-$M$ classical vector into $\log_2(M)$ qubits (\textit{i.e.} amplitude encoding), and used it to demonstrate that several classes of real-world classical data can be loaded in short depth. Further, we demonstrated AMLET's superior performance over a ``layer-by-layer'' algorithm that was inspired by previous work. The methods we introduced are depth-tunable, \textit{i.e.} one can trivially achieve a trade-off between accuracy and depth. The benefits of AMLET include: (a) depth-accuracy tunability, (b) absence of any optimization parameters (eliminating concerns about optimization time and/or barren plateaus), (c) absence of ancilla qubits, and (d) shorter-depth results than the layer-by-layer approach.
Our numerical studies showed data being loaded in orders of magnitude lower depth than the general loading algorithm would yield. 
This study suggests paths to valuable future work: for instance, one could apply AMLET to encodings more complex than the standard amplitude encoding; another possibility is a modified version of AMLET that adheres to an arbitrary circuit pattern or hardware topology. 
In conclusion, besides introducing a new algorithm that will be useful in quantum circuit compilation, this work demonstrates that the ``input problem'' will likely \textit{not} be an obstacle to quantum advantage for many scientifically interesting classical workloads.

\section*{Acknowledgements}

We are grateful to Shavindra Premaratne, Roza Kotlyar, and Anne Matsuura for useful discussions, and we thank Daan Camps for providing valuable feedback on the manuscript.

\bibliography{refs}

\end{document}